# Blackbody radiation and distribution function with three parameters

Liu Changshi

Physics division, Department of mechanical and electrical engineering, Jiaxing College, Zhejiang, 314001, P. R. China

Telephone: 086-0575-88067065, fax: 086-0575-88319253, E-mail:liucs4976@sohu.com

## Abstract

The object of this work is to give the key to answer whether or not there is another numerical method which is different from the equation proposed by Planck to predict blackbody radiation by frequency. Firstly, Maxwell distribution function for molecule velocities was modified, resulting in a distribution function with three parameters for deriving monochromatic intensity of blackbody radiation through frequency. Then this simulation function was applied to estimate the energy density (Jm$^{-2}$) of blackbody radiation by frequency at 5500 K, 5000 K, 4500 K, 4000 K, 3500 K and 2.73 K of temperature. The results of density simulated by means of distribution function suggested in this paper agree well with the experimental data. All of the correlation coefficients between actual and calculated data are 1.0; at the same time, the mean relative errors are less than 6.65% in total.

1. Introduction

The influence of blackbody radiation has been great on both theory and technology since physicists begun to explore such radiation. Based on the Wien`s displacement relationship [1], Planck`s equation to describes the distribution of radiation emitted by a blackbody [2], the features of blackbody radiation have been applied in many fields, such as, cosmic microwave background radiation [3], temperature measurements of astrophysical object, color temperature, infrared temperature, combustion formed by laser induced incandescence for soot particle measurements and two-color pirometry. This kind of radiation also plays a prominent role in the physics of climate [4,5].

Because Planck introduced an element of energy which can be considered as the initiation quantum physics, there has been research on Planck`s radiation law since Planck published his paper in 1901. There are various ways to derive Planck`s radiation law [6-12] except for Planck`s original one. Although the introduction of energy quanta a century ago has led to the currently accepted explanation within quantum theory, there is still no firm conclusion as to whether or not blackbody radiation can be predicted within numerical method. In this paper, in according with the comparison between the experimental curves of monochromatic intensity emitted by blackbody and curve of distribution function for molecule velocities derived by Maxwell, modification of the latter results in a distribution function with three parameters by which the relationship between frequency and monochromatic intensity of blackbody radiation is expressed in mathematics way. The experimental values agree well with calculated results given by the distribution function described above.

2. Simulation method

The measured behaviors of monochromatic intensity (energy per unit area) radiated by a blackbody, for a given frequency $v$ and at absolute temperature of 5500 K, 5000 K, 4500 K, 4000 K and 3500 K are summarized in figure 1, respectively.

There is one peak for each curve. By tracking $v_{max}$ (the frequency corresponding to the peak of energy density in $Jm^{-2}$), the $v_{max}$ in the monochromatic intensity $\rho(v)$ at 3500 K is around $3.623 \times 10^{14}$ Hz, it is about $5.70 \times 10^{14}$ Hz for 5500 K. This characteristic means that energy density radiated by a blackbody per unit area increase rapidly in the range $o \prec v \prec v_{max}$, whereas the drop of them is in slow way when the frequency is higher than $v_{max}$. Therefore, every curve is a cusp, especially for monochromatic intensity of blackbody at 3500 K of temperature. However, if shape of each curve of monochromatic intensity is in comparison with the shape of distribution function for molecule velocities derived by Maxwell, the form of monochromatic intensity curve is in the form of distribution function likely. Enlightenment was from the distribution function for molecule velocities derived by Maxwell, this function is written as

$$\rho(v) = 4\pi \left(\frac{m}{2\pi kT}\right)^{\frac{3}{2}} \cdot v^2 \cdot e^{-\frac{m}{2kT} \cdot v^2} \quad (1)$$

It is reasonable to express this function in another form

$$\rho(v) = p_1 \cdot v^{p_2} \cdot e^{-p_3 \cdot v^{p_4}} \quad (2)$$

Where, $p_1 = 4\pi \left(\frac{m}{2\pi kT}\right)^{\frac{3}{2}}$ appears as one parameter, $p_2 = 2$ is another parameter, and $p_3 = \frac{m}{2kT}$ stands for the third last parameter and the last parameter $p_4$ equals to $p_2$. It is known that the curve of distribution function for molecule velocities derived by Maxwell is symmetrical, but the curves of monochromatic intensity radiated by a blackbody as function of frequency are asymmetrical. It can be sure that as long as some modification is carried out on Maxwell's distribution function, one function will be able to be obtained for describing the relationship between frequency and energy density in $Jm^{-2}$. The strategies for this perfection as follows, parameter $p_1$ is kept in function (2) firstly, then parameter $p_2$ which is power of independence variable will be changeable so that the monochromatic intensity can increase rapidly along frequency in the range of $o \prec v \prec v_{max}$, in order to maintain the drops of energy density are slow, in the step three, parameter $p_3$ is changeable too and the parameter $p_4$ which is power of independence variable in exponential function appears in 1. Of course, independence variable is frequency $v$. Based on these considers, one distribution function with three parameters was written as below

$$\rho(v) = p_1 \cdot v^{p_2} \cdot e^{-p_3 v} \quad (3)$$

Of course, the first parameter $p_1$, the second parameter $p_2$, and the last parameter $p_3$ are to be determined through fitting.

3. Simulation

This paper analyzed the emission spectra of blackbody under different conditions of temperature according to the above mentioned approach (3). Then through measured raw data being employed in method of regression analysis, the functions for simulation of monochromatic intensity from frequency were obtained and listed in table 1. Figure 2 compares the monochromatic intensity radiated by a blackbody calculated by the present models in table 1 at very higher temperature with the measured ones. Spectrum of cosmic microwave background radiation simulated using the present models in table 1 is also compared with experimental ones in figure 3 [13,14].

The correlation coefficient between measured and calculated data is given in table 1, the magnitude of each correlation coefficient is 1.0. The mean relative error (ARE) ($\frac{1}{n}\sum_{i=1}^{i=n}\frac{|\rho_{io}-\rho_{is}|}{\rho_{io}}\times 100\%$) for evaluation simulation results is also shown in table 1. The maximum value of ARE is 6.55%, the maximum data of relative error ($\frac{|\rho_{io}-\rho_{is}|}{\rho_{io}}$) takes place at the lowest frequency for every spectrum. As it is seen from table 1, figure 2 and 3, there is a good agreement between calculated monochromatic intensity emitted from blackbody by functions in table 1 and measured ones according to the spectra of background radiation.

4. Conclusions

In order to predict monochromatic intensity of blackbody radiation by means of method, which is different from the equation derived by Planck, a distribution function with three parameters has been developed by revision on distribution function for molecule velocities derived by Maxwell. The curve of the energy density radiated by a blackbody in per unit area calculated from numerical simulation is fitted to the experimental data against frequency. The theoretical values obtained by a nonlinear parameter estimation technique are well consistent with experimental data. The good agreement between measured and calculated results of monochromatic intensity emitted by blackbody indicates that the proposed model can predict some aspects of blackbody radiation accurately; all of these are helpful for our understanding of what behavior of blackbody emission is in mathematical method. Also following the spirit of current work, it is expected to be relative straightforward to extend this model in other subjects.

Figure caption

Figure 1 Spectra of blackbody radiation (frequency vs. intensity) for different temperatures.

Figure 2 Comparisons between experimental and calculated blackbody radiation (frequency vs. intensity) at different temperature

Figure 3 Comparisons between experimental and calculated spectrum of cosmic microwave background radiation (frequency vs. intensity) at temperature of 2.73 K

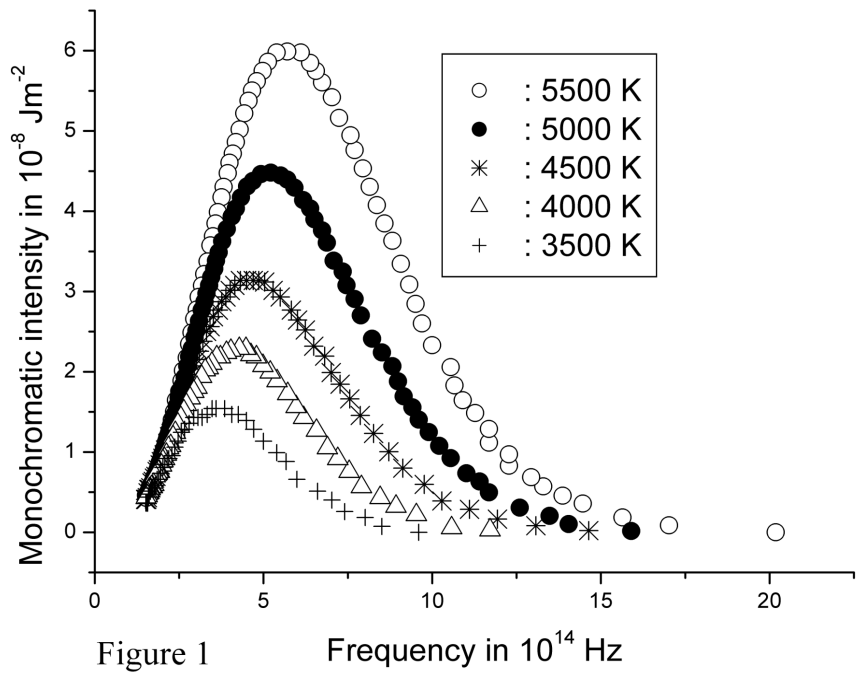

Figure 1

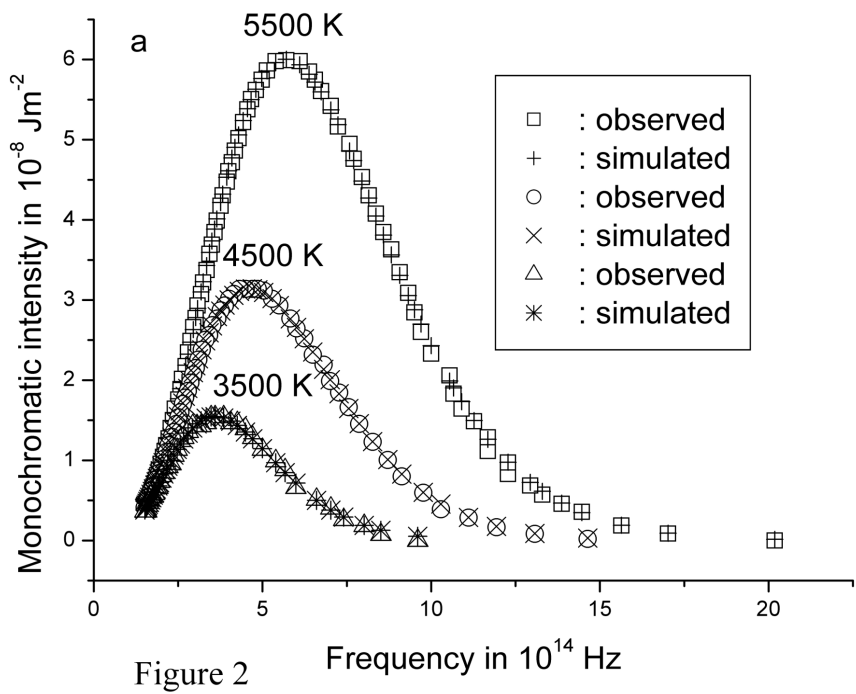

Figure 2

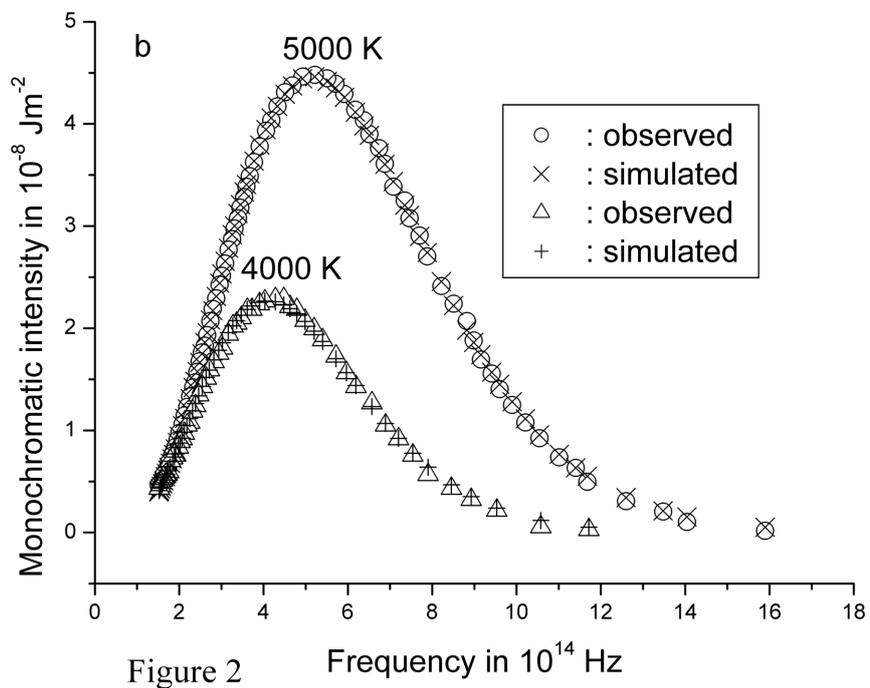

Figure 2

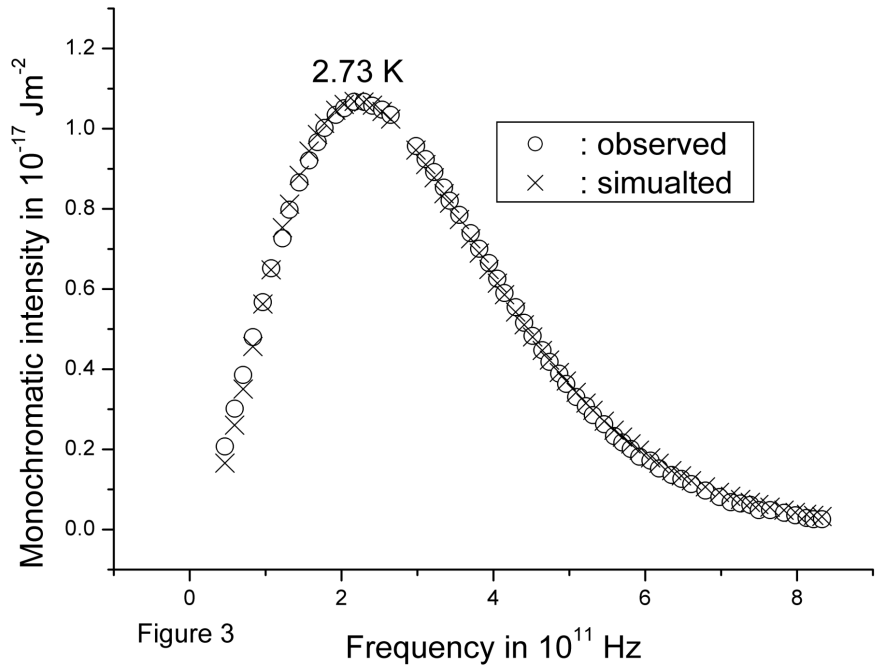

Figure 3

Table 1. Results and judge of simulation for the monochromatic intensity of blackbody radiation

| Temperature K | Function form | Correlation Coefficient | Mean Relative Error |
|---|---|---|---|
| 5500 | $\rho(v) = 4.0 \times 10^{-75} v^{4.69} e^{-8.2 \times 10^{-15} v}$ | 1.0 | 2.70% |
| 5000 | $\rho(v) = 9.13 \times 10^{-75} v^{4.67} e^{-9.0 \times 10^{-15} v}$ | 1.0 | 5.55% |
| 4500 | $\rho(v) = 2.17 \times 10^{-73} v^{4.58} e^{-9.9 \times 10^{-15} v}$ | 1.0 | 6.55% |
| 4000 | $\rho(v) = 2.74 \times 10^{-75} v^{4.72} e^{-1.15 \times 10^{-14} v}$ | 1.0 | 6.31% |
| 3500 | $\rho(v) = 3.68 \times 10^{-78} v^{4.93} e^{-1.37 \times 10^{-14} v}$ | 1.0 | 5.0% |
| 2.75 | $\rho(v) = 4.27 \times 10^{-44} v^{2.42} e^{-1.1 \times 10^{-11} v}$ | 1.0 | 6.63% |